\documentclass{article}
\usepackage{amsmath,graphicx}
\usepackage[utf8]{inputenc}
\usepackage{lipsum}% http://ctan.org/pkg/lipsum
\usepackage{amssymb}
\usepackage{multicol}
\usepackage{float}
\usepackage{footnote}
\usepackage{url}
\DeclareMathOperator{\E}{\mathbb{E}}
\usepackage{mathrsfs}
\newcommand{\ve}[1]{\textbf{#1}}		% vectors
\newcommand{\di}[1]{\mathcal{#1}}		% distributions

\usepackage[preprint]{spconf}
\copyrightnotice{\copyright\ IEEE 2019}
\toappear{To appear in {\it Proc.\ ICASSP2019, May 12-17, 2019, Brighton, UK}}

\usepackage[dvipsnames]{xcolor}

\title{\textit{Wav2Pix}: Speech-conditioned Face Generation using Generative Adversarial Networks}
\name{Amanda Duarte\(^1\)\(^2\), Francisco Roldan\(^1\), Miquel Tubau\(^1\), Janna Escur\(^1\), Santiago Pascual\(^1\)}
\secondlinename{Amaia Salvador\(^1\), Eva Mohedano\(^3\), Kevin McGuinness\(^3\), Jordi Torres\(^1\)\(^2\), Xavier Giro-i-Nieto\(^1\)\(^2\)}
\address{\(^1\)Universitat Polit\`ecnica de Catalunya, Barcelona, Catalunya/Spain \\
\(^2\)Barcelona Supercomputing Center, Spain \\
\(^3\)Insight Centre for Data Analytics - DCU, Ireland}

\begin{document}
\ninept
\maketitle
\begin{abstract}
%The abstract should contain about 100 to 150 words.
%Image synthesis have been a challenging task for the AI community in recent years.
%-----------------------------------------------------------------------
Speech is a rich biometric signal that contains information about the identity, gender and emotional state of the speaker.
In this work, we explore its potential to generate face images of a speaker by conditioning a Generative Adversarial Network (GAN) with raw speech input. 
We propose a deep neural network that is trained from scratch in an end-to-end fashion, generating a face directly from the raw speech waveform without any additional identity information (e.g reference image or one-hot encoding). 
Our model is trained in a self-supervised approach by exploiting the audio and visual signals naturally aligned in videos. With the purpose of training from video data, we present a novel dataset collected for this work, with high-quality videos of youtubers with notable expressiveness in both the speech and visual signals.
%\footnote{To appear at 44th International Conference on Acoustics, Speech, and Signal Processing, ICASSP 2019.}

%-----------------------------------------------------------------------
\end{abstract}
\section{Introduction}
% \amanda{Needs to be shorter}
% -------------------------------------------------
% MOdel Diagram 
%---------------------------------------------------
\begin{figure*}[!ht]
\centering
\includegraphics[width=\textwidth]{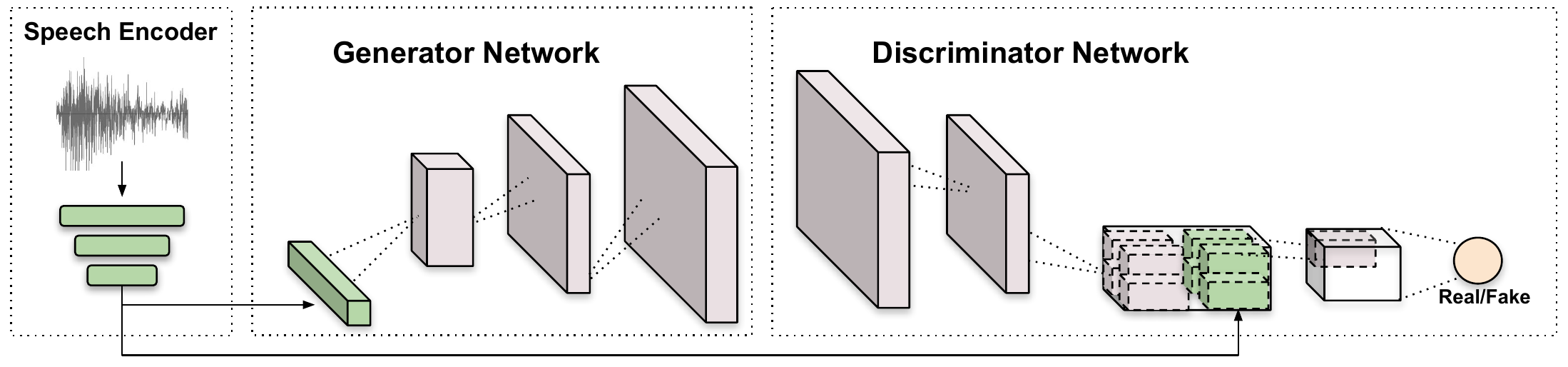}
\caption{The overall diagram of our speech-conditioned face generation GAN architecture. The network consists of a speech encoder, a generator and a discriminator network. An audio embedding (green) is used by both the generator and discriminator, but its error is just back-propagated at the generator. It is encoded and projected to a lower dimension (vector of size 128). Pink blocks represent convolutional/deconvolutional stages.}
\label{fig:architecture}
\end{figure*}
%-----------------------------------------------------------------

% ----------------------------------------------------------------
% General info and motivation 
Audio and visual signals are the most common modalities used by humans to identify other humans and sense their emotional state.
Features extracted from these two signals are often highly correlated, allowing us to imagine the visual appearance of a person just by listening to their voice, or build some expectations about the tone or pitch of their voice just by looking at a picture of the speaker.
When it comes to image generation, however, this multimodal correlation is still under-explored. 
%  Challenges
%Going one step further from cross-modal retrieval or zero-shot learning that typically rely on learning joint representations. Generative models further require modelling details of the samples and learning a complex generative function that produces meaningful outputs for images and/or sounds never been seen or heard before.

% Amaia: The above commented paragraph is unnecessary in my opinion.

%Present our idea
In this paper, we focus on cross-modal visual generation, more specifically, the generation of facial images given a speech signal. 
% Difference from popular works
Two recent approaches have recently popularized this research venue~\cite{said_that, obama}.
Chung~\textit{et al.}~\cite{said_that} present a method for generating a video of a talking face starting from audio features
%the Mel-frequency cepstral coefficients (MFCC) of a speech frame of the target speaker 
and an image of him/her (identity).
Suwajanakorn \textit{et al.} focus on animating a point-based lip model to later synthesize high quality videos of President Barack Obama~\cite{obama}. 
Unlike the aforementioned works, however, we aim to generate the whole face image at pixel level, conditioning only on the raw speech signal (\emph{i.e.} without the use of any handcrafted features) and without requiring any previous knowledge (e.g speaker image or face model). 

% Dataset
To this end, we propose a conditional generative adversarial model (shown in Figure \ref{fig:architecture}) that is trained using the aligned audio and video channels in a self-supervised way. 
For learning such a model, high quality, aligned samples are required. 
This makes the most commonly used datasets such as \textit{Lip Reading in the wild}~\cite{lrw}, or \textit{VoxCeleb}~\cite{voxceleb} unsuitable for our approach, as the position of the speaker, the background, and the quality of the videos and the acoustic signal can vary significantly across different samples.
We therefore built a new video dataset from YouTube, composed of videos uploaded to the platform by well-established users (commonly known as \textit{youtubers}), who recorded themselves speaking in front of the camera in their personal home studios. 
Such videos are usually of high quality, with the faces of the subject featured in a prominent way and with notable expressiveness in both the speech and face.
% Contribution
Hence, our main contributions can be summarized as follows:

1) We present a conditional GAN that is able to generate face images directly from the raw speech signal, which we call \emph{Wav2Pix}.
% and without providing any identity information. (e.g gender or emotion) 
    
2) We present a manually curated dataset of videos from youtubers, that contains high-quality data with notable \emph{expressiveness} in both the speech and face signals.

3) We show 
%evaluate our approach on our datasets 
%including the commonly used VoxCeleb~\cite{voxceleb} and further show qualitatively 
that our approach is able to generate realistic and diverse faces.% given a speech signal. 

The developed model, software and dataset are publicly released\footnote{\url{https://imatge-upc.github.io/wav2pix/}}.

\label{sec:intro}

\section{Related works}
% ----------------------------------------------------------
% General explanation about GAN, Contitioned GANs and LSGAN
%-----------------------------------------------------------

\noindent\textbf{Generative Adversarial Networks}. (GANs)~\cite{gan} are a state of the art deep generative model that consist of two networks, a Generator $G$ and a Discriminator $D$, playing a min-max game against each other. This means both networks are optimized to fulfill their own objective: $G$ has to generate realistic samples and $D$ has to be good at rejecting $G$ samples and accepting real ones. This joint learning adversarial process lasts for as long as $G$ begins generating samples which are as good enough as to fool $D$ into making as many mistakes as possible. The way Generator can create novel data mimicking real one is by mapping samples $z \in \mathbb{R}^{n}$ of arbitrary dimensions coming from some simple prior distribution $\di{Z}$ to samples $x$ from the real data distribution $\di{X}$ (in this case we work with images, so $\ve{x}\in\mathbb{R}^{w\times h\times c}$ where $w\times h$ are spatial dimensions width and height and $c$ is the amount of channels).This means each $\ve{z}$ forward is like sampling from $\di{X}$. On the other hand the discriminator is typically a binary classifier as it distinguishes~\textit{real} samples from~\textit{fake} ones generated by $G$. One can further condition $G$ and $D$ on a variable $e\in\mathbb{R}^{k}$ of arbitrary dimensions to derive the the conditional GANs~\cite{cGAN} formulation, with the conditioning variable being of any type, \emph{e.g.} a class label or text captions~\cite{textoimage}. In our work, we generate images conditioned on raw speech waveforms.

Numerous improvements to the GANs methodology have been presented lately. Many focusing on stabilizing the training process and enhance the quality of the generated samples~\cite{photographicGAN, high_fidelity}. Others aim to tackle the vanishing gradients problem due to the sigmoid activation and the $\log$-loss in the end of the classifier~\cite{wGan, began, LSGAN}. To solve this, the least-squares GAN (LSGAN) approach \cite{LSGAN} proposed to use a least-squares function with binary coding (1 for real, 0 for fake). We thus use this conditional GAN variant with the objective function is given by:  

\begin{equation}
\begin{aligned}
  \underset{D}\min~V_{\text{LSGAN}}(D) & = \frac{1}{2}\E_{\ve{x},\ve{e}\sim p_{\text{data}}(\ve{x}, \ve{e})}[(D(\ve{x},\ve{e}) - 1)^{2}]\\
    &+ \frac{1}{2}\E_{\ve{z}\sim p_{\ve{z}}(\ve{z}),\ve{e}\sim p_{\text{data}}(\ve{e})}[D(G(\ve{z},\ve{e}),\ve{e})^{2}].
\label{eq:lsgan_d}
\end{aligned}
\end{equation}
\begin{equation}
\begin{aligned}
\underset{G}\min~V_{\text{LSGAN}}(G) = \frac{1}{2}\E_{\ve{z}\sim p_{\ve{e}}(\ve{e}),\ve{y}\sim p_{\text{data}}(\ve{y})}[(D(G(\ve{z},\ve{e}),\ve{e}) - 1)^{2}],
\label{eq:lsgan_g}
\end{aligned}
\end{equation}

\textbf{Multi-modal generation}. Data generation across modalities is becoming increasingly popular~\cite{DCGAN, textoimage, ACGAN, stackgan}. Several works~\cite{textoimage, stackgan} present different approaches for synthesizing realistic images given a text description. Recently, a number of approaches combining audio and vision have appeared, with tasks such as generating speech from a video~\cite{speech_recons} or generating images from audio/speech~\cite{s2i}. In this paper we will focus on the latter.

Most works on audio conditioned image generation adopt non end-to-end approaches and exploit previous knowledge about the data. Typically, speech has been encoded with handcrafted features
%, such us the MEL spectrum or Mel-frequency Cepstral Coefficients (MFCC), 
which have been very well engineered to represent human speech. At the visual part, point-based models of the face \cite{karras2017audio} or the lips \cite{obama} have been adopted. In contrast to that, our network is trained entirely end-to-end solely from raw speech to generate image pixels. %, and neither hand-crafted features nor additional pre-training is used.

A direct synthesis of facial pixels was obtained in~\cite{said_that} with a discriminative model whose input were a pair of audio features and a visual example of the face to predict. In that case, the model had the help of a additional identity information (image of the speaker) to help in the prediction, so the network learned how to modify this input to match with the speech utterance. Following a similar architecture, a generative model trained with adversarial training was proposed in~\cite{e2espeech}. In this case, they introduced a temporal regularization to improve the smoothness of the output video sequence. Our work differs from theirs in that we use raw speech instead of hand-crafted features, and we do not need any image of the speaker as all identity information is extracted from the speech only. %The main difference of our network with these two works is its lack of the visual identity as an input, which is learned and encoded in the network parameters during training and the usage of raw speech signal instead of hand-crafted audio features extracted before hand.

\label{sec:related_work}

\section{Youtubers Dataset}
\label{sec:dataset_pipeline}
%The lack of appropriate datasets are one of the issues for the progress in the development of cross-modal generative models specially in the case of speech-to-images generation. 
%For the creation of our dataset, we took advantage of the audiovisual content 
In this section we describe the multi-stage pipeline adopted to collect the new audio-visual dataset of human speech used in this work.
We collected videos uploaded to YouTube by well-established users (so-called \textit{youtubers}), who tend to record themselves speaking in front of the camera in a well controlled environment. 
Such videos are usually of high quality, with the faces of the subject featured in a prominent way and with notable expressiveness in both the speech and face.
The Youtubers dataset is composed of two sets: the complete noisy dataset automatically generated, and a clean subset which was manually curated to obtain high quality data.

\begin{figure}
\centering
\includegraphics[width=\columnwidth]{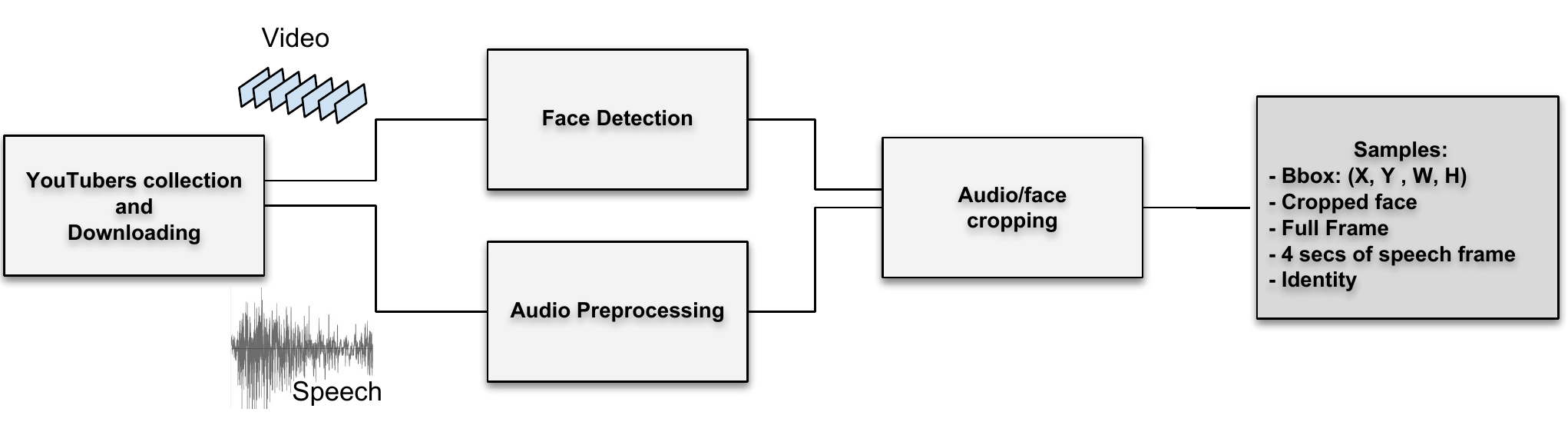}
\caption{High level representation of the data collection pipeline. Each detected face is associated with a 4 seconds length audio and the corresponding identity. Besides that we also kept the bounding box coordinates and the original image frame.}
\label{fig:dataset_pipeline}
\smallskip
\end{figure}

%\subsection{Full dataset}
In total we collected 168,796 seconds of speech with the corresponding video frames, and cropped faces from a list of 62 youtubers active during the past few years. The dataset is gender balanced and manually cleaned keeping 42,199 faces, each with an associated 1-second speech chunk.
The pipeline used for downloading and pre-processing the full dataset is summarized in Figure \ref{fig:dataset_pipeline}, and the key stages are discussed in the following paragraphs:

%\begin{table}[!h]
%\begin{center}
%\caption{Summary of the dataset making a breakdown based on the ID of the speaker. Resulting dataset contains approximately \amanda{x} hours of speech.}
%\label{table:dataset}
%\begin{tabular}{llll}
%\hline\noalign{\smallskip}
%Sex & Speakers & \hspace{0.2cm}Faces & \hspace{0.2cm}Speech (sec)\\
%\noalign{\smallskip}
%\hline
%\noalign{\smallskip}
%Male  & \hspace{0.25cm} 29 & \hspace{0.2cm}26299 &  \hspace{0.5cm} 105196\\
%Female & \hspace{0.25cm} 33 & \hspace{0.2cm}15900 & \hspace{0.5cm}  63600\\
%\hline
%\textbf{TOTAL} & \hspace{0.25cm} 62 & \hspace{0.12cm} 42199 & \hspace{0.5cm}  %168796\\
%\end{tabular}
%\end{center}
%\end{table}
%\setlength{\tabcolsep}{1.4pt}

\textbf{YouTubers collection and downloading:} A list of 62 different Spanish speaker \textit{youtubers} was built, consisting on 29 males and 33 females from different ethnicity and accents.
%This list was chosen accordantly to their popularity and expressions in front of the camera. After having the final list, the last 15 videos uploaded to their channel were downloaded. 

\textbf{Audio preprocessing:} The audio was originally downloaded in Advance Audio Coding (AAC) format at 44100 Hz and stereo and converted to WAV, as well as re-sampled to 16 kHz with 16 bits per sample and converted to mono.

\textbf{Face Detection:} The faces were detected using a Haar Feature-based Classifier \cite{haar} trained with frontal face features. We prevent the method from having false positives by taking only the most confident detection for each frame. 

\textbf{Audio/faces cropping:} From each detection it is saved the bounding box coordinates, an image of the cropped face in BGR format, the full frame and a 4 seconds length speech frame, which encompasses 2 seconds ahead and behind the given frame. Moreover, we keep an identity (name) for each sample. We apply a pre-emphasis step to each speech frame and normalize it between $[-1, 1]$. %to increase the amplitude of the higher frequency bands while decreasing the amplitude of the lower ones., as higher frequencies are more important for signal disambiguation.

% A post cleaning was made in a subset of 10 classes of our dataset in order to verify our premise of having a cleaned dataset, e.g. no false positives in the face detection and a clean acoustic in the speech signal, would improve the synthesis of faces or not. 
% A data augmentation was made in order to obtain a larger number of data samples. As state before, the original data contain a 4 seconds lenght of audio. To perform the data augmentation process, we duplicate each image by 5 times and for each correspondent audio file we applied a random point (? better word) choice of a center and choose half size from the left and half size from the right to crop 1 second of speech signal. In result of that we were able to have ..... training samples. (to upsample our training data in 5 times)

As stated in section \ref{sec:experiments} our model demonstrate a loss of performance when trained with noisy data. Thus, a part of the dataset was manually filtered to obtain the high-quality data required to improve the performance of our network.
We took a subset of 10 identities, five female and five male, from our dataset and manually filtered them making sure that all faces were visually clear and all audios contain speech, so that all the silence and music parts were removed. As a result, the cleaned dataset contains a total of 4,860 images and audios (4 seconds length). 

%Further experiments were perform using the VoxCeleb dataset \cite{voxceleb}. VoxCeleb is an audio-visual dataset consisting of short clips of human speech, extracted from interview videos uploaded to YouTube. \amanda{This experiments show that we were not able to produce faces using this dataset. We believe that it is because the audios are pretty dirty. It shows that our dataset is better but also shows that our method is not that robust... not sure how to show this in here.}

% ---------------------------------------------------------------
% Results Image
%---------------------------------------------
\begin{figure*}[!ht]
\centering
\includegraphics[width=0.95\textwidth]{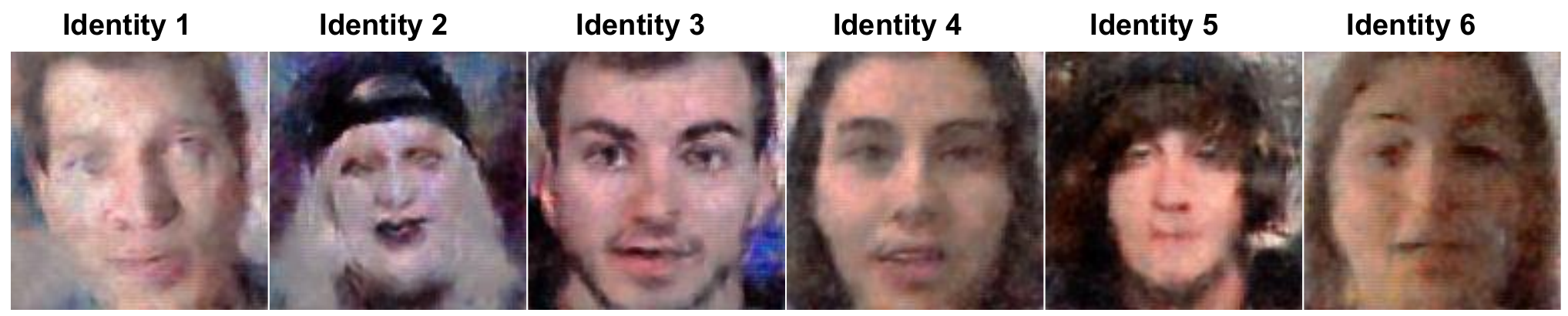}
\caption{Generated samples conditioned to raw speech produced by our model.}
\label{fig:resultsidentity}
\end{figure*}

%\begin{figure}
%\centering
%\includegraphics[width=\columnwidth]{dataset_examples.pdf}
%\caption{Dataset examples \amanda{expressions}}
%\label{fig:dataset_pipeline}
%\smallskip
%\end{figure}

\label{sec:training}

\section{Method}
% -------------------------------------------------
% General Info
%---------------------------------------------------
Since our goal is to train a GAN conditioned on raw speech waveforms, our model is divided in three modules trained altogether end-to-end: a speech encoder, a generator network and a discriminator network described in the following paragraphs respectively.
The speech encoder was adopted from the discriminator in \cite{segan}, while both the image generator and discriminator architectures were inspired by \cite{textoimage}. The whole system was trained following a Least Squares GAN~\cite{LSGAN} scheme. Figure~\ref{fig:architecture} depicts the overall architecture.

% -------------------------------------------------
% Speech Encoder
%---------------------------------------------------
\textbf{Speech Encoder:} 
As mentioned in Section \ref{sec:related_work}, many existing methods~\cite{said_that, e2espeech, obama} require the extraction of handcrafted audio features  before feeding the data into the neural network. This could limit the representation learning, as the audio information is extracted manually and not optimized for our generative task. In contrast, SEGAN \cite{segan} proposed a method for speech enhancement in which they do not work on the spectral domain, but at the waveform level. We coupled a modified version of the SEGAN discriminator $\Phi$ as input to an image generator $G$. Our speech encoder was modified to have $6$ strided one-dimensional convolutional layers of kernel size $15$, each one with stride $4$ followed by LeakyReLU activations. Moreover we only require one input channel, so our input signal is $\ve{s}\in\mathbf{R}^{T\times 1}$, being $T=16,384$ the amount of waveform samples we inject into the model (roughly one second of speech at $16\,kHz$). The aforementioned convolutional stack decimates this signal by a factor $4^6 = 4096$ while increasing the feature channels up to $1024$. Thus, obtaining a tensor $f(\ve{s}) \in \mathbb{R}^{4\times 1024}$ in the output of the convolutional stack $f$. This is flattened and injected into three fully connected layers that reduce the final speech embedding dimensions from $1024\times 4 = 4096$ to $128$, obtaining the vector $\ve{e} = \Phi(\ve{s}) \in\mathbb{R}^{128}$. 

% -------------------------------------------------
% Generator Network
%---------------------------------------------------
\textbf{Image Generator Network:} 
We take the speech embedding $\ve{e}$ as input to generate images such that $\hat{\ve{x}} = G(\ve{e}) = G(\Phi(\ve{s}))$. The inference proceeds with two-dimensional transposed convolutions, where the input is a tensor $\ve{e}\in\mathbb{R}^{1\times 1\times 128}$ (an image of size $1\times 1$ and $128$ channels), based on~\cite{DCGAN}. The final interpolation can either be $64\times 64\times 3$ or $128\times 128\times 3$ just by playing with the amount of transposed convolutions ($4$ or $5$). It is important to mention that we have no latent variable $\ve{z}$ in $G$ inference as it did not give much variance in predictions in preliminary experiments. To enforce the generative capacity of $G$ we followed a dropout strategy at inference time inspired by~\cite{isola2017image}.

In preliminary experiments, we found it convenient to add a secondary component to the loss of $G$: a \textit{softmax} classifier trained over the given speech embedding. This classifier helped the whole network into preserving the identity of the speaker. The magnitude of the classification component is controlled by a new hyper-parameter $\lambda$. Therefore, the $G$ loss, follows the LSGAN loss presented in Equation~\ref{eq:lsgan_g} with the addition of this weighted auxiliary loss for identity classification.
%, which becomes: 

%%\begin{equation}
%\begin{aligned}

%\underset{G}\min~V_{\text{LSGAN}}(G) = \frac{1}{2}\E_{\ve{z}\sim p_{\ve{z}}(\ve{z}),\ve{y}\sim p_{\text{data}}(\ve{y})}[(D(G(\ve{z},\ve{y}),\ve{y}) - 1)^{2}] + \\
%\lambda.
  
%\label{eq:lsgan_g}
%\end{aligned}
%\end{equation}

% -------------------------------------------------
% Discriminator Network
%---------------------------------------------------
\textbf{Image Discriminator Network:} The Discriminator $D$ is designed to process several layers of stride 2 convolution with a kernel size of 4 followed by a spectral normalization \cite{snorm} and leakyReLU (apart from the last layer). When the spatial dimension of the discriminator is $4\times 4$, we replicate the speech embedding $\ve{e}$ spatially and perform a depth concatenation. The last convolution is performed with stride 1 to obtain a $D$ score as the output.

%---------------------------------------------------

\label{sec:architecture}

\section{Experiments}
\label{sec:experiments}
%\amanda{Main questions to be discussed here:}

%1: Does the higher resolution in this case helps? 128x128 instead of 64x64
%--- Yes, but in order to train with higher resolution we need more data.
%-- Best results are using 1s, 128x128 Youtubers-v2 with data augmentation (we copy each each and audio 5 times, let the image the same and resample the audio randomly cutting the 4 in 5 different ones of 1s)

%2: Adding a classifier on top of the speech embedding helps the model to learn the identity of the speaker and generate a better face for a given speaker?
%* NOPE, the architecture still not learn the identity... 

%3: Besides the classifier on top of the speech embedding, adding the ID information to the generator helps ?
%* YES for the image quality but not to learn the identity

%4: In terms of dataset, what is best in this case?
%Large and dirty (v1 and VoxCeleb) - very bad, compare the results of voxCeleb with ours
%Small and clean (V2) - too little for traing the network, not enough information to generate images 
%Large and clean (V2-DA) -- this one!

%5: Different speech length as input: 
%4s, 1s, 300ms
%* 1s is enough if we have enough data... our conclusions were wrong... new results prove it
%Problem: All the fid store are pretty high. The best fid results do not agree with the
%best images generated. The one that got better score are very bad images. Not sure how to show this in the paper. No quantitative results are ok?

%trained with:...}

\textbf{Model training:} The \textit{Wav2Pix} model was trained on the cleaned dataset described in Section \ref{sec:dataset_pipeline} combined with a data augmentation strategy. % to overcome the limited amount of data.
In particular, we copied each image five times, pairing it with 5 different audio chunks of 1 second randomly sampled from the 4 seconds segment.
Thus, we obtained $\approx$ 24k images and paired audio chunks of 1 second used for training our model.
Our implementation is based on the PyTorch library \cite{pyTorch} and trained on a GeForce Titan X GPU with 12GB memory. 
We kept the hyper-parameters as suggested in \cite{textoimage}, changing the learning rate to 0.0001 in G and 0.0004 in D as suggested in~\cite{heusel2017gans}. We use ADAM solver \cite{adam} with momentum 0.1.

\begin{figure}[!h]
    \centering
    \includegraphics[width=0.98\linewidth]{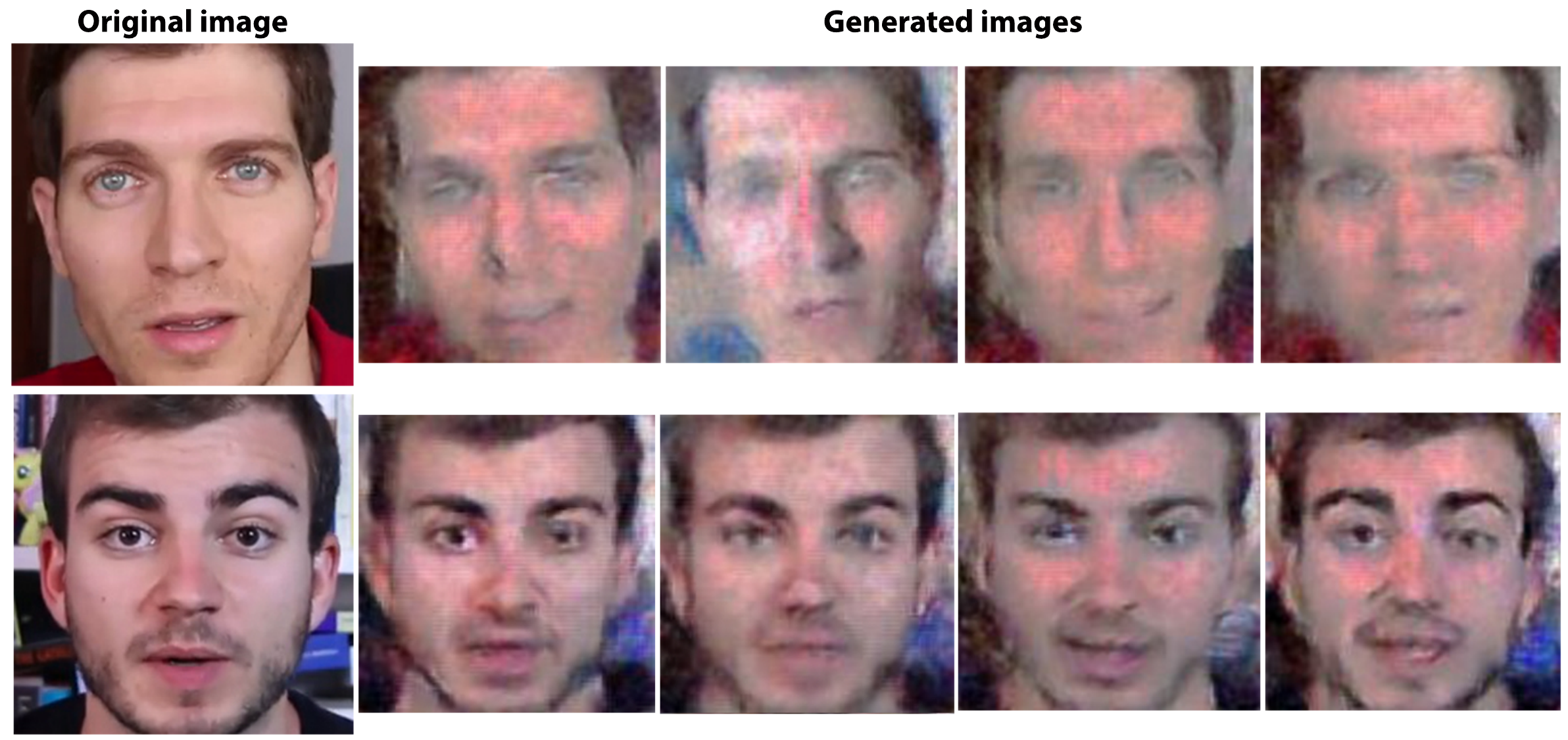}
    \caption{Examples of generated faces compared to the original image of the person who the voice belongs to. In the generated images, we can observe that our model is able to preserve the physical characteristics and produce different face expressions. In the first row we can see examples of the \textit{youtuber Javier Muñiz}. In the second row we can see examples of the \textit{youtuber Jaime Altozano}.}
    \label{fig:original_generated}
\end{figure}

\textbf{Evaluation: } 
Figure \ref{fig:original_generated} shows examples of generated images given a raw speech chunk, compared to the original image of the person who the voice belongs to. Different speech waveform produced by the same speaker were fed into the network to produce such images.
Although the generated images are blurry, it is possible to observe that the model learns the person's physical characteristics, preserving the identity, and present different face expressions depending on the input speech \footnote{Some examples of images and it correspondent speech as well as more generated images are available at: https://imatge-upc.github.io/wav2pix/}. 
Other examples from six different identities are presented in Figure \ref{fig:resultsidentity}.

% Face Classification metric
To quantify the model's accuracy regarding the identity preservation, we fine-tuned a pre-trained VGG-Face Descriptor network \cite{deep_face_recognition, VGGFace_pretrained_dataset} with our dataset. We predicted the speaker identity from the generated images of both the speech train and test partitions, obtaining an identification accuracy of 76.81\% and 50.08\%, respectively.

% We also assessed the ability of the model for generating realistic faces, independently from the preservation of the identity.
% In this case we ran the well known Viola \& Jones face detector \cite{landmarks_paper} over the generated images. 
% We were able to identify the 68 key-points of a face in 90.25\% of the images generated from our speech test set. 

\begin{figure}[H]
    \centering
    \includegraphics[width=0.98\linewidth]{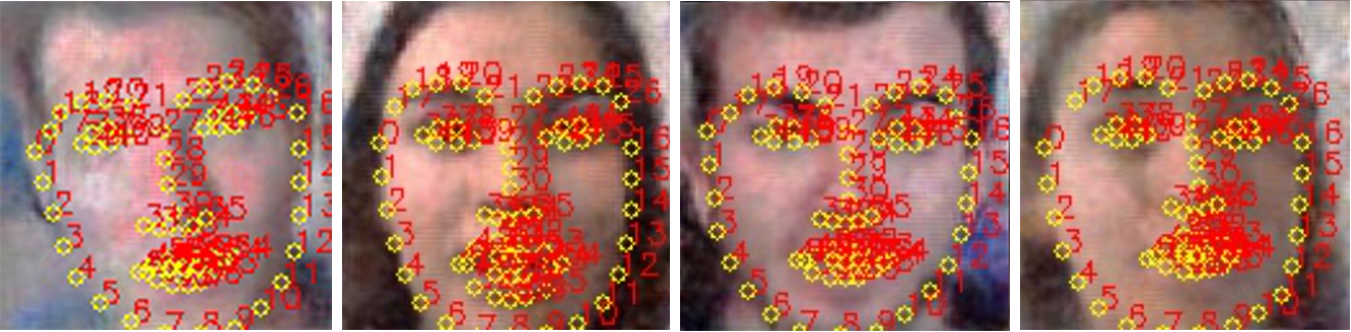}
    \caption{Examples of the 68 key-points detected on images generated by our model. Yellow circles indicate facial landmarks fitted to the generated faces, numbered in red fonts.
    }
    \label{fig:ladmarks}
\end{figure}

We also assessed the ability of the model to generate realistic faces, regardless of the true speaker identity. To have a more rigorous test than a simple Viola \& Jones face detector~\cite{haar}, we measured the ability of an automatic algorithm \cite{landmarks} to correctly identify facial landmarks on images generated by our model. We define detection accuracy as the percentage of images where the algorithm is able to identify \textit{all} 68 key-points. For the proposed model and all images generated for our test set, the detection accuracy is 90.25\%, showing that in most cases the generated images retain the basic visual characteristics of a face. 
This detection rate is much higher than the identification accuracy of 50.08\%, as in some cases the model confuses identities, or mixes some of them in a single face. 
Examples of detected faces together with their numbered facial landmarks can be seen in Figure \ref{fig:ladmarks}.

\textbf{Additional experiments: }
We also tried to generate faces with noisy speech, experiments that resulted in failures.
Firstly, we used audio snippets from the same yotubers videos that presented background noise, silences or other people's voice.
Secondly, using the well known VoxCeleb 1 dataset \cite{voxceleb}, which contains a larger amount of images and identities but present a lower audio quality.
In both cases, the quality of results was very poor, making it almost impossible to recognize faces.
These results show the importance of having clean speech samples to train the proposed model.

\begin{figure}[!h]
    \centering
    \includegraphics[width=0.98\linewidth]{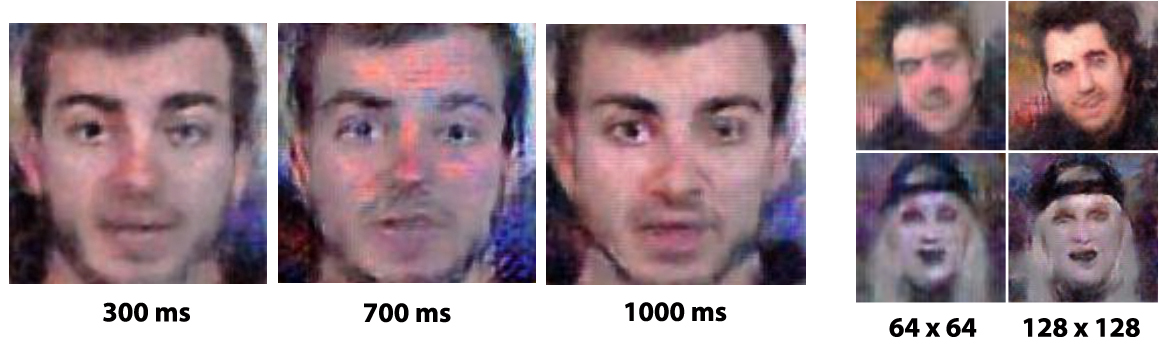}
    \caption{(Left) Images generated for three speech chunk lengths. (Right) Images generated at two spatial resolutions.}
    \label{fig:scales}
\end{figure}

We also observed a drop in performance when working with smaller speech chunks and lower image definitions. We observed a visual degradation when using audio chunks of 300 and 700 milliseconds, which was reflected in a decrease of the face detection rate. 
%The detection ratio was 81.16\% when using 300 ms and 89.12\% when using 700 ms, both of them worse than the previously 90.25\% when using 1000 ms.
Detection accuracy when using 300 and 700 ms chunks was 81.16\% and 89.12\%, respectively, in both cases worse than the 90.25\% accuracy achieved when using 1000 ms chunks.
Figure \ref{fig:scales} (left) shows examples of generated images for the three speech chunk lengths. 
Figure \ref{fig:scales} (right) shows how using a lower definition of 64x64 pixels instead of 128x128 results into blurrier images.

\label{sec:results}

\section{Conclusions}

In this work we introduced a simple yet effective cross-modal approach for generating images of faces given only a short segment of speech, and proposed a novel generative adversarial network variant that is conditioned on the raw speech signal. 

As high-quality training data are required for this task, we further collected and curated a new dataset, the Youtubers dataset, that contains high quality visual and speech signals. Our experimental validation demonstrates that the proposed approach is able to synthesize plausible facial images with an accuracy of 90.25\%, while also being able to preserve the identity of the speaker about 50\% of the times. Our ablation experiments further showed the sensitivity of the model to the spatial dimensions of the images, the duration of the speech chunks and, more importantly, on the quality of the training data. 

Further steps may address the generation of a sequence of video frames aligned with the conditioning speech, as well exploring the behaviour of the \textit{Wav2Pix} when conditioned on unseen identities.

% In this work we presented a simple but yet effective first cross-modal approach for generating images of faces using generative adversarial networks conditioned on raw speech signal.
% % We demonstrated that our model can synthesize plausible facial images in 90.25\% of the images associated or not to the identity of the speaker. 
% Our experiments also showed the sensitivity of the model to the spatial definition of the images, the duration of the speech chunks and, more importantly, on the quality of the training data. As a result, the Youtubers dataset we collected automatically needed to be manually curated to obtain high quality visual and speech data.
% Further steps may address the generation of a sequence of video frames aligned with the conditioning speech, as well exploring the behaviour of the \textit{Wav2Pix} when conditioned by unseen identities.

\section*{Acknowledgements}
%Amanda:
This research has received funding from “la Caixa” Foundation funded by the European Union’s Horizon 2020 research and innovation programme under the Marie Skłodowska-Curie grant agreement No. 713673.
%The research leading to these results has received funding from “la Caixa” Foundation funded by the European Union’s Horizon 2020 research and innovation programme under the Marie Skłodowska-Curie grant agreement No. 713673.
This research was partially supported by MINECO and the ERDF under contracts TEC 2015-69266-P and TEC 2016-75976-R, by the MICINN under the contract TIN 2015-65316-P and the Programa Severo Ochoa under the contract SEV-2015-0493. 
This research was also partially supported by SFI under grant number SFI/15/SIRG/3283.
We gratefully acknowledge the support of NVIDIA Corporation with the donation of GPUs.
We would like to thank the youtubers \textit{Javier Muñiz and Jaime Altozano} for providing us the rights of publishing their images on this research.

%The research leading to these results has received funding from “la Caixa” Foundation. This project has received funding from the European Union’s Horizon 2020 research and innovation programme under the Marie Skłodowska-Curie grant agreement No. 713673.
\label{sec:conclusions}

\bibliographystyle{IEEEbib}
\bibliography{refs}

\end{document}